# Study of tantalum and iridium as adhesion layers for Pt/LGS high temperature SAW devices

Thierry AUBERT, Omar ELMAZRIA, Badreddine ASSOUAR, Laurent BOUVOT, Zoumnone BOURNEBE, Michel HEHN, Sylvain WEBER, Mourad OUDICH, Patrick ALNOT
Institut Jean Lamour
UMR 7198 CNRS – Nancy University
54506 Vandoeuvre lès Nancy, France
thierry.aubert@lpmi.uhp-nancy.fr

*Abstract*—In this paper, we report on the use of tantalum and iridium as adhesion layers for platinum electrodes used in high temperature SAW devices based on langasite substrates (LGS). Unlike iridium, tantalum exhibits a great adhesive strength, and a very low mobility through the Pt film, ensuring a device lifetime of at least half an hour at 1000°C. The latter is limited by morphological modifications of platinum, starting by the apparition of crystallites on the surface, and followed by important terracing and breaking of the film continuity. SNMS and XRD measurements allowed us to show that these phenomena are likely intrinsic to platinum film, whatever be the nature of the adhesion layer. Finally, after having outlined a possible scenario leading to this deterioration, we consider some solutions that could replace platinum in order to increase the lifetime of LGS-based SAW devices in high temperatures conditions.

*Keywords-high temperature; platinum; SAW; langasite; tantalum*

## I. Introduction

It is now well established that SAW devices can be used as pressure, gas or temperature sensors. Those devices are particularly interesting for applications in harsh environments where no electronics can survive. Indeed, because they can be remotely powered and interrogated, embedded electronic is not required for wireless applications. Many industrial sectors, such as aeronautics, metallurgy, automotive, are looking forward those small, robust and wireless devices [1,2]. Efforts are now focused on selecting materials able to withstand harsh conditions, such as high temperatures up to 1000°C.

Regarding to the substrate, several studies have shown that langasite ($La_3Ga_5SiO_{14}$) is well suited for this kind of applications [3]. It is only limited by its relative important propagation losses, especially at high frequencies, making it unsuitable for applications at high temperatures above 1 GHz [4], which could be an embarrassing limitation for wireless applications. The challenge now is to find a reliable composition of materials to realize IDT structure with high stability and good performances.

Up to now, most of the studies have been conducted on platinum electrodes [1,2]. Indeed, this noble metal has a relatively high melting temperature (1773°C), and an important chemical inertness appreciable at high temperature. Nevertheless, this interesting property has also a drawback : the consequence of the platinum weak chemical affinity for others elements is a very poor adhesion of platinum films on oxide substrates such as quartz or langasite. The solution consists in using a very thin (10 to 20 nm) intermediate layer between the substrate and the platinum film, so-called "adhesion layer". It must be constituted by a refractive metal, more reactive than platinum, such as titanium, zirconium, tantalum … Bonding between the adhesion layer and the platinum film is ensured by alloying [5].

Several studies have shown the limitations of titanium as adhesion layer at temperatures above 600°C [6]. The diffusion of titanium in the platinum film is actually large, leading to the formation of oxide precipitates ($TiO_2$) in the Pt grain boundaries. The consequences are the progressive vanishing of the adhesion layer leading to delamination, the modification of the electrical properties of the IDTs and even their destruction.

Pt/Zr combination has shown far better performances until 700°C. In particular, AES (Auger electron spectroscopy) measurements have highlighted that Zr has not diffused at all through Pt after 8h of annealing at 700°C [7]. Nonetheless, it has been shown that this combination is inadequate for continuous operation beyond about 700°C, because of de-wetting phenomena [3].

More recently, Da Cunha *et al.* have studied alternative solutions, replacing platinum by a Pt/10%Rh alloy, while maintaining the adhesion layer of zirconium [3]. This combination exhibited a very good stability until 950°C, which is a considerable advance. This stability was even improved by co-depositing the alloy with $ZrO_2$, and by using a SiAlON protective over-layer [8].

Our purpose in this work is to study new adhesion layers for platinum, not studied yet in high temperature SAW applications, to reach better results than the ones with zirconium, and then to ensure a device lifetime of at least 30 minutes at 1000°C. We will focus our efforts on two potential candidates : tantalum and iridium.

Tantalum as adhesion layer for platinum has already been studied in other fields than SAW applications. Maeder *et al.*, in particular, have noted that its performance is very close to that of zirconium after an annealing of 10 minutes in a low pressure $O_2$ atmosphere at 620°C [9]. On the other hand, iridium is a noble metal, just like platinum. Nevertheless, a thin oxide film forms on its surface, even at room temperature, foreshadowing a better chemical reactivity than that of platinum, which could allow bonding with the substrate [10].

## II. EXPERIMENTAL

Platinum and tantalum thin films were deposited on YX langasite substrates by sputtering method, while iridium was evaporated in electron beam system. The thickness of the adhesion layers was systematically 10 nm, while that of platinum film was 100 nm. SAW delay lines operating at 167 MHz were then processed combining photolithography and ion beam etching (IBE). All devices were then electrically characterized with a probe station before and after annealing. These were performed in a tube furnace under air atmosphere at temperatures from 900°C to 1000°C and for a duration varying between 30 minutes and 2 hours. The samples were quickly inserted in the pre-heated furnace, and pulled out of the hot zone as fast as possible. Langasite substrates showed great resistance to this thermal shock treatment.

The morphology of the thin metal films was studied by X-ray diffraction (XRD) and scanning electron microscopy (SEM). Second neutral mass spectroscopy (SNMS) was used to obtain their compositional depth profiles.

## III. RESULTS

### A. Pt/Ir Combination

If the adhesion strength of the as-deposited Pt/Ir films seemed good, the lithography process revealed that it was actually poor. In fact, after this phase, approximately 70 % of the achieved devices were unusable, because of the presence of numerous scratches on the IDTs. We can suppose that their origin is related to a poor adhesion of the Pt/Ir film on the substrate, making the devices particularly delicate. This fragility also appeared when the prober tips come into contact with the IDT pads during electrical characterization. Significant scratches were then systematically observed. This behavior was confirmed after the high temperature annealing. SEM images revealed the formation of numerous blisters as well as delamination, leading to a strong thinning of the IDTs fingers, threatening their continuity (see Fig. 1). However, the IDTs have not been completely broken, as the $S_{21}$ response attests (see Fig. 2). One can guess that the result would have been probably different if the experiment had lasted over half an hour. This strong lack of adhesion, partially expected due to the noble character of iridium, leads us to focus our efforts on the Pt/Ta combination.

### B. Pt/Ta Combination

Unlike Pt/Ir combination, Pt/Ta films showed very good adhesion to the substrate at room temperature, but also after annealing. The frequency response shows no deterioration after annealing at 1000°C for 30 minutes (see Fig. 3). SEM images confirmed this result, showing only a starting modification, with the apparition of crystallites at the surface of the platinum film (see Fig. 4).

To study the limitations of this kind of device at high temperature, we made it undergo a new annealing process at 900°C for 2 hours. The result was no appeal : the SAW signal vanished while SEM images revealed that the crystallites have disappeared, leaving a substantial terracing and breaking of the IDTs fingers, which explains the loss of the SAW signal (see Fig. 5).

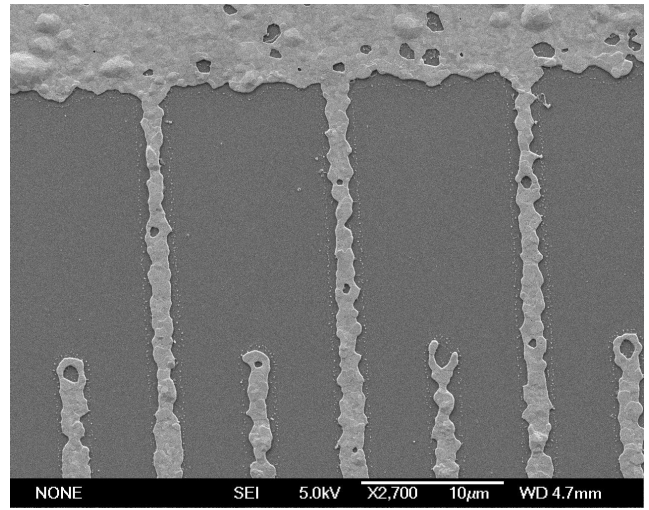

Figure 1. SEM image of a Pt/Ir IDT after annealing for 30 min. at 1000°C

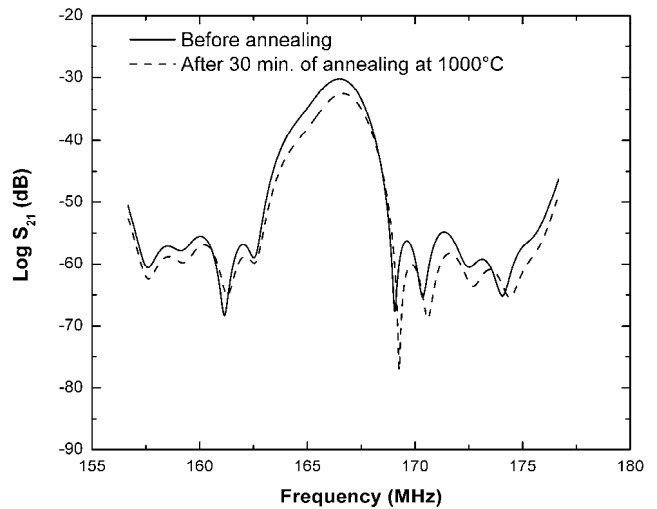

Figure 2. Frequency response of a Pt/Ir/LGS SAW device before and after annealing for 30 min. at 1000°C

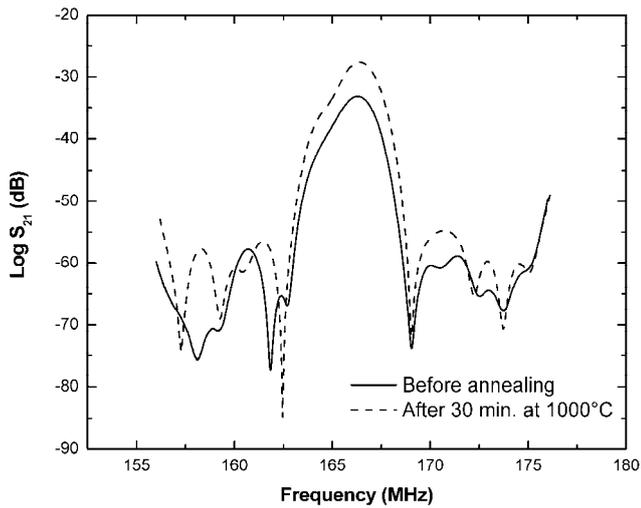

Figure 3. Frequency response of a Pt/Ta/ LGS SAW device before and after annealing for 30 min. at 1000°C

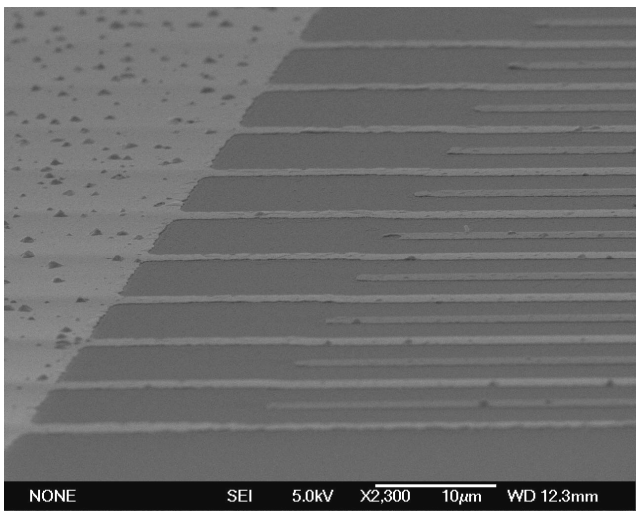

Figure 4. Grazing SEM image of a Pt/Ta IDT before and after annealing for 30 min. at 1000°C

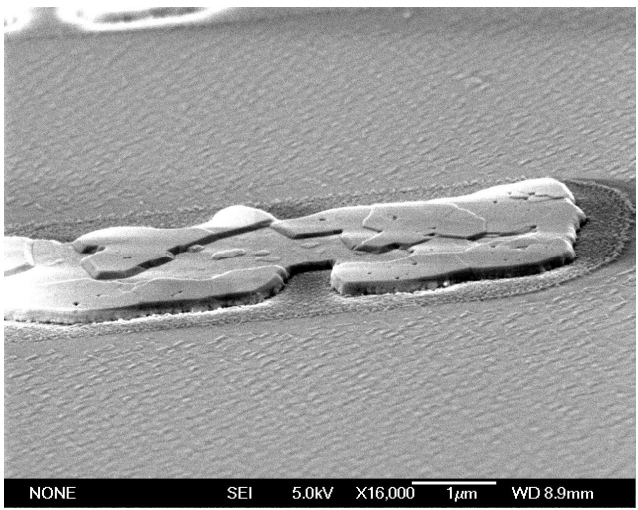

Figure 5. Grazing SEM image of a Pt/Ta IDT finger tip after 2 hours. of annealing at 900°C

In order to increase the lifetime of this kind of devices at very high temperatures, it is essential to understand the origin of the crystallites appearing at the surface of the platinum film during annealing. Indeed, it seems clear that they constitute a prerequisite for the deterioration of the IDTs. That is why some unprocessed Pt/Ta/LGS samples were then investigated by SNMS (Fig. 6) and XRD (Fig. 7).

## IV. Discussion

On SNMS depth profile curve, the atomic concentration of several elements is plotted versus the time of abrasion, which is related to the depth inside the sample. The two vertical lines symbolize the approximate interfaces Pt/Ta and Ta/LGS. One can see that there has not been massive diffusion of tantalum through the Pt film during annealing, and a slight one into the LGS substrate. This result, coupled with the excellent adhesive strength of the Pt/Ta films, shows how tantalum is a good adhesion layer for platinum thin films and for high temperature applications. Moreover, tantalum will probably play no role in the development of the crystallites on the surface of the Pt layer, implying that this phenomenon is intrinsic to platinum. XRD measurements allow confirming this hypothesis. θ-2θ scan pointed out a strong [111] orientation of the platinum films, which is a conventional result for FCC metals, like Pt [11]. Another interesting information is given by the shift of this [111] Pt peak if one compares its position before and after annealing. The peak of as-deposited films is at a slightly lower angle of diffraction (39,538°) than its theoretical value (39,763°), which means that the interplanar value $d_{111}$ of these films is larger than the bulk one. This indicates that the Pt layer is slightly in compression in the plane of the film. A likely reason for this compressive stress is peening, by which the subsurface of a sputtered film is compressed due to bombardment by reflected neutral atoms [11]. To the opposite, the peak after annealing has shifted to a much higher value (39,850°). This can be explained by the difference of the thermal expansion coefficient α between Pt ($8,8.10^{-6}$ $K^{-1}$) and LGS ($5,6.10^{-6}$ $K^{-1}$), which induces, after cooling down at a high rate a strain of the film and an associated "thermal" stress. Here, as $α_{LGS}$ is lower than $α_{Pt}$, we can deduce that the Pt film is stretched in the plane, implying a lower $d_{111}$ value than for the bulk, and thus a higher diffraction angle than the theoretical value. More important, it can be clearly seen on this spectrum that the full width at half maximum (FWHM) of this [111] Pt peak decreases significantly while its intensity increases after annealing (Fig. 7). It is well known that the width of diffraction peaks decreases while increasing crystallinity, so we can conclude that the Pt layer strongly recrystallized during annealing, and that the apparition of the crystallites is a harbinger of this phenomenon. This scenario can also explain the mechanism leading to the loss of continuity of the Pt film after long-time annealing. Indeed, if the temperature or the time of annealing increases, the phenomenon of recrystallization goes on, and crystals grow up out of plane more and more, which implies that some areas of the film will be depleted of Pt in order to "feed" these baby-crystals.

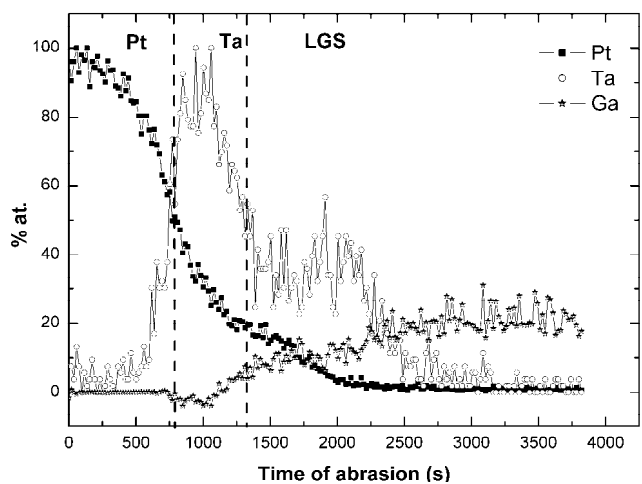

Figure 6. SNMS depth profile of Pt/Ta/LGS sample after 30 min. of annealing at 1000°C

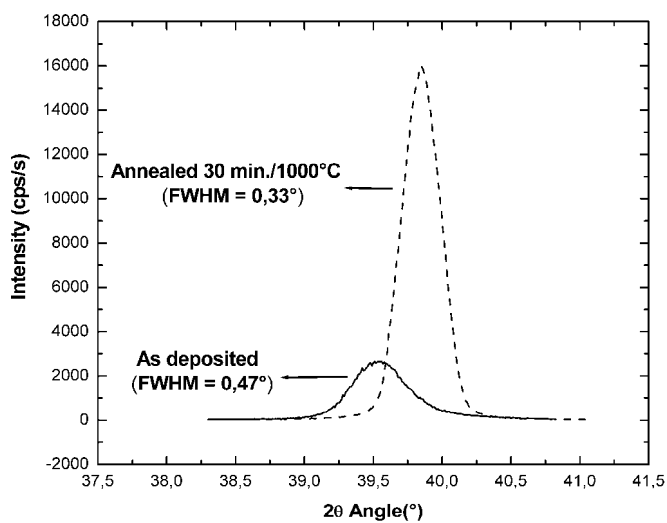

Figure 7. Pt [111] XR diffraction peak before and after annealing for 30 min. at 1000°C

This interpretation means that, whatever be the nature of the adhesion layer, the deterioration of the platinum film at high temperature cannot be avoided, because of this recrystallization phenomena. This is consistent with the fact that similar observations of Pt film recrystallization during annealing have been made in the case of a zirconium adhesion layer [3].

As our metal thin films are polycrystalline, grain boundaries are very numerous and the total surface of grains is large, which implies that interface diffusion phenomena are probably predominant under bulk ones, and govern the recrystallization process. Moreover, the onset of surface diffusion is about the Hüttig temperature (which is defined as $0{,}3T_m$ with $T_m$ the melting temperature in K, i.e. for Pt about 340°C), whereas lattice mobility becomes significant over the Tammann temperature (defined as $0{,}5T_m$, i.e. for Pt about 750°C) [11]. However, it is clear that the higher is the melting temperature, the latter the recrystallisation phenomena will start. A trivial solution to increase the lifetime of the device would be to replace platinum by a noble metal having a significantly higher melting temperature, i.e. ruthenium or iridium. Nonetheless, they start oxidizing at temperatures about 700°C, and their oxides are highly volatile [12]. Another interesting way is that followed by Da Cunha *et al.* who use a Pt/10%Rh alloy. Indeed, in theses kind of alloys, precipitates form in the grain boundaries, which prevents grain boundaries diffusion [13].

V. CONCLUSION

To summarize, we have shown that tantalum has the required properties to be used as an adhesion layer for platinum IDTs. The limitation of 30 min at 1000 °C is attributed to platinum and not to the layer of tantalum. The deterioration of Pt/Ta electrode is explained by a strong recrystallization of platinum as evidenced by the results of X-ray diffraction.